\documentclass{ws-procs9x6}

\usepackage{amssymb,amsmath,amsfonts}
\usepackage{graphicx}
\usepackage{graphics}
\usepackage{eepic,epsfig}

\begin{document}

\title{TOPOLOGICAL CASIMIR EFFECT IN \\ NANOTUBES AND NANOLOOPES}

\author{A. A. SAHARIAN$^*$}

\address{Department of Physics, Yerevan State University,\\
1 Alex Manoogian Street, 0025 Yerevan, Armenia\\
$^*$E-mail: saharian@ictp.it}

\begin{abstract}
The Casimir effect is investigated in cylindrical and toroidal
carbon nanotubes within the framework of the Dirac-like model for
the electronic states. The topological Casimir energy is positive
for metallic cylindrical nanotubes and is negative for
semiconducting ones. The toroidal compactification of a
cylindrical nanotube along its axis increases the Casimir energy
for metallic-type (periodic) boundary conditions along its axis
and decreases the Casimir energy for the semiconducting-type
compactifications. For finite length metallic nanotubes the
Casimir forces acting on the tube edges are always attractive,
whereas for semiconducting-type ones they are attractive for small
lengths of the nanotube and repulsive for large lengths.
\end{abstract}


\bodymatter

\section{Introduction}

An interesting application of the quantum field theoretical models
with non-trivial topology recently appeared in nanophysics. The
long-wavelength description of the electronic states in graphene
can be formulated in terms of the Dirac-like theory with the Fermi
velocity playing the role of speed of light (see, e.g., Ref.
\cite{Cast09}). Single-walled carbon nanotubes are generated by
rolling up a graphene sheet to form a cylinder and the
background spacetime for the corresponding Dirac-like theory has topology $%
R^{2}\times S^{1}$. Compactifying the direction along the cylinder axis we
obtain another class of graphene made structures called toroidal carbon
nanotubes with the background topology $R^{1}\times (S^{1})^{2}$.

The boundary conditions imposed on the fermionic field in these
nanostructures give rise to the modification of the spectrum for
vacuum fluctuations and, as a result, to the Casimir-type
contributions in vacuum expectation value (VEV) of the
energy-momentum tensor (for the topological Casimir effect see
\cite{Most97}). In the present paper, based on Refs.
\cite{Bell09a,Bell09b}, we investigate the fermionic Casimir
effect in nanotubes.  The paper is organized as follows. In the
next section the topological Casimir effect is considered in
cylindrical nanotubes. The VEV of the energy-momentum tensor for a
fermionic field in toroidal nanotubes is discussed in Sec.
\ref{sec:Tor}. The Casimir forces acting on the edges of
finite-length carbon nanotubes are investigated in Sec.
\ref{sec:FinLength}. The main results are summarized in Sec.
\ref{sec:Conclusion}.

\section{Cylindrical nanotubes}

\label{sec:Cyl}

A single wall cylindrical nanotube is a graphene sheet rolled into
a cylindrical shape. For this case we have spatial topology
$R^{1}\times S^{1}$ with the compactified dimension of the length
$L$. The carbon nanotube is characterized by its chiral vector
$\mathbf{C}_{h}=n_{w}\mathbf{a }_{1}+m_{w}\mathbf{a}_{2}$, with
$n_{w}$, $m_{w}$ being integers, and
$L=|\mathbf{C}_{h}|=a\sqrt{n_{w}^{2}+m_{w}^{2}+n_{w}m_{w}}$, $a=|\mathbf{a}_{1}|=|%
\mathbf{a}_{2}|=2.46\mathring{A}$. A zigzag nanotube corresponds
to the special case $\mathbf{C}_{h}=(n_{w},0)$, and a armchair
nanotube corresponds to the case $\mathbf{C}_{h}=(n_{w},n_{w})$.
All other
cases correspond to chiral nanotubes. In the case $%
n_{w}-m_{w}=3q_{w}$, $q_{w}\in Z$, the nanotube will be metallic
and in the case $n_{w}-m_{w}\neq 3q_{w}$ the nanotube will be
semiconductor with an energy gap inversely proportional to the
diameter.

The electronic band structure of a graphene sheet close to the
Dirac points shows a conical dispersion
$E(\mathbf{k})=v_{F}|\mathbf{k}|$, where $\mathbf{k}$ is the
momentum measured relatively to the Dirac points and $v_{F}$
represents the Fermi velocity. The corresponding low-energy
excitations can be described by a pair of two-component Weyl
spinors. By taking into account that in the presence of an
external magnetic field an effective mass term is generated for
the fermionic excitations, we consider the general case of massive
spinor field $\psi $ on background of $3$-dimensional flat
spacetime with spatial topology $R^{1}\times S^{1}$. The
corresponding line
element has the form $ds^{2}=dt^{2}-(dz^{1})^{2}-(dz^{2})^{2}$, where $%
-\infty <z^{1}<\infty $ and $0\leqslant z^{2}\leqslant L$.

We assume that the field obeys the boundary condition
\begin{equation}
\psi (t,z^{1},z^{2}+L)=e^{i\beta }\psi (t,z^{1},z^{2}),  \label{BCcyl}
\end{equation}%
with a constant phase $\beta $. For metallic nanotubes $\beta =0$
and for semiconductor nanotubes, depending on the chiral vector,
we have two classes of inequivalent boundary conditions
corresponding to $\beta =\pm 2\pi /3$. In the expression for the
Casimir densities the phase $\beta $ appears in the form $\cos
(n\beta )$ and, hence, the Casimir energy density and stresses are
the same for these two cases. The VEV of the energy-momentum
tensor for a cylindrical nanotube has the form (no summation over
$l$)
\begin{equation}
\langle T_{l}^{k}\rangle _{\mathrm{(cyl)}}(L)=\delta
_{l}^{k}\sum_{n=1}^{\infty }\cos (n\beta )G^{(l)}(nmL)\frac{e^{-nmL}}{\pi
L^{3}n^{3}},  \label{EMTNano}
\end{equation}%
with the notations $G^{(0)}(z)=G^{(1)}(z)=1+z$,
$G^{(2)}(z)=-(2+2z+z^{2})$. In particular, the Casimir energy
density is positive for metallic nanotubes and negative for
semiconducting ones.

In the massless case we have $\langle T_{0}^{0}\rangle
_{\mathrm{(cyl)}}=\langle T_{1}^{1}\rangle
_{\mathrm{(cyl)}}=-\langle T_{2}^{2}\rangle
_{\mathrm{(cyl)}}/2=S_{\beta }/(\pi L^{3})$, with the notation
$S_{\beta }=\sum_{n=1}^{\infty }\cos (n\beta )/n^{3}$. In
particular, $S_{0}=1.202$ and $S_{2\pi /3}=-0.534$. In carbon
nanotubes we have two sublattices and each of them gives the
contribution to the Casimir densities given by (\ref{EMTNano}).
So, for the Casimir energy density on a carbon
nanotube with radius $L$ one has $\langle T_{0}^{0}\rangle _{1,1}^{\mathrm{%
(cn)}}=2\hbar v_{F}S_{\beta }/(\pi L^{3})$, where the standard units are
restored.

\section{Toroidal nanotubes}

\label{sec:Tor}

For the geometry of a toroidal nanotube we have the spatial topology $%
(S^{1})^{2}$. The boundary conditions have the
form%
\begin{equation}
\psi (t,z^{1}+L_{1},z^{2})=e^{i\beta _{1}}\psi
(t,z^{1},z^{2}),\;\psi (t,z^{1},z^{2}+L_{2})=e^{i\beta _{2}}\psi
(t,z^{1},z^{2}).  \label{BCTor}
\end{equation}%
The corresponding energy density and the vacuum stresses are given by
the expressions (no summation over $l$)%
\begin{eqnarray}
\langle T_{0}^{0}\rangle _{\mathrm{(tor)}} &=&\sum_{j=1,2}\langle
T_{0}^{0}\rangle _{\mathrm{(cyl)}}(L_{j})+\frac{2}{\pi }\sum_{m_{1}=1}^{%
\infty }\sum_{m_{2}=1}^{+\infty }\frac{1+mg(\mathbf{L}_{2},\mathbf{m}_{2})}{%
g^{3}(\mathbf{L}_{2},\mathbf{m}_{2})} \nonumber  \\
&&\times \frac{\cos (m_{1}\beta _{1})\cos (m_{2}\beta _{2})}{\exp (mg(%
\mathbf{L}_{2},\mathbf{m}_{2}))}, \label{T00D2tor}\\
\langle T_{l}^{l}\rangle _{\mathrm{(tor)}} &=&\langle T_{0}^{0}\rangle _{%
\mathrm{(tor)}}-\frac{m^{5}}{\pi }\sum_{j=1,2}\sum_{m_{j}=1}^{+\infty }\cos
(m_{j}\beta _{j})L_{l}^{2}m_{l}^{2}G(mL_{j}m_{j}) \notag \\
&-&\frac{2m^{5}}{\pi }\sum_{m_{1},m_{2}=1}^{+\infty }\cos
(m_{1}\beta _{1})\cos (m_{2}\beta
_{2})L_{l}^{2}m_{l}^{2}G(mg(\mathbf{L}_{2},\mathbf{m}_{2})),
\label{TllD2tor}
\end{eqnarray}%
with $G(x)=(3+3x+x^2)e^{-x}/x^5$, $l=1,2$, and $g(\mathbf{L}_{2},\mathbf{m}_{2})=\sqrt{%
m_{1}^{2}L_{1}^{2}+m_{2}^{2}L_{2}^{2}}$. The corresponding
formulae for the Casimir densities in toroidal nanotubes are
obtained with an additional factor 2. In standard units the factor
$\hbar v_{F}$ appears as well. From the formulae given above it
follows that the toroidal compactification of a cylindrical
nanotube along its axis increases the Casimir energy for periodic
conditions ($\beta _{1}=0$) and decreases the Casimir energy for
the semiconducting-type compactifications.

\section{Finite-length nanotubes}

\label{sec:FinLength}

In this section we will assume that the nanotube has finite length
$a$. We assume the periodicity condition (\ref{BCcyl}) along the
compact dimension. As the Dirac field lives on the cylinder
surface it is natural to impose bag boundary conditions
$(1+i\gamma ^{l}n_{l})\psi =0$, $z^{1}=0,a$, on the cylinder
edges, with $\gamma ^{l}$ being the Dirac matrices and $n_{l}$ is
the normal to the boundaries. The additional confinement of the
electrons along the tube axis leads to the change of the ground
state energy. The corresponding
Casimir energy is decomposed as%
\begin{equation}
E=aL\langle T_{0}^{0}\rangle
_{\mathrm{(cyl)}}+2E^{(1)}-\frac{1}{\pi }\sum_{l=-\infty
}^{+\infty }\int_{m_{l}}^{\infty
}\frac{xdx}{\sqrt{x^{2}-m_{l}^{2}}}\ln \left(
1+\frac{x-m}{x+m}e^{-2ax}\right) , \label{Edecomp}
\end{equation}%
with $m_{l}^{2}=[(2\pi l+\beta )/L]^{2}+m^{2}$. The part $E^{(1)}$
is the Casimir energy for a single edge (when the other edge is
absent) in the half-space. The last term in Eq. (\ref{Edecomp}) is
the interaction part. The single edge part of the Casimir energy
does not depend on the length of the tube and will not contribute
to the Casimir force.

\begin{figure}[tbph]
\begin{center}
\epsfig{figure=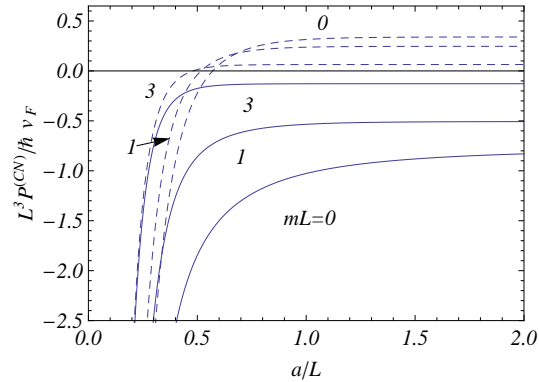,width=7cm,height=5.cm}
\end{center}
\caption{The fermionic Casimir forces acting on the edges of the
metallic (full curves) and semiconducting-type (dashed curves)
nanotubes as functions of the tube length.} \label{fig}
\end{figure}

For the Casimir force acting on the edges of the tube we have%
\begin{equation}
P=-\langle T_{0}^{0}\rangle _{\mathrm{(cyl)}}-\frac{2}{\pi
L}\sum_{l=-\infty }^{+\infty }\int_{m_{l}}^{\infty
}dx\,\frac{x}{\frac{x+m}{x-m}e^{2ax}+1}.  \label{PCN}
\end{equation}
The corresponding expressions for the Casimir energy and force in
finite length cylindrical nanotubes are obtained with an
additional factor 2. So, for the Casimir force acting per
unit length of the edge of a carbon nanotube one has: $P^{\mathrm{(CN)}%
}=2\hbar v_{\mathrm{F}}P$. For long tubes, $a/L\gg 1$, the first
term on the right of (\ref{PCN}) is dominant and we have
$P^{\mathrm{(CN)}}\approx
-0.765\hbar v_{\mathrm{F}}/L^{3}$ for metallic nanotubes and $P^{\mathrm{(CN)%
}}\approx 0.34\hbar v_{\mathrm{F}}/L^{3}$ for semiconducting ones.
In the limit $a/L\ll 1$ the Casimir force do not depend on the
chirality and one has $P^{\mathrm{(CN)}}\approx -0.144\hbar
v_{\mathrm{F}}/a^{3}$. In Fig. \ref{fig} we have plotted the
Casimir forces acting on the edges of metallic (full curves) and
semiconducting-type (dashed curves) carbon nanotube as functions
of the tube length for different values of the fermion mass. As it
is seen, for metallic nanotubes these forces are always
attractive, whereas for semiconducting-type ones they are
attractive for small lengths and repulsive for large lengths.

\section{Conclusion}

\label{sec:Conclusion}

We have investigated the Casimir effect for cylindrical and
toroidal nanotubes within the framework of the Dirac-like model
for electrons. The VEV of the energy-momentum tensor is given by
formula (\ref{EMTNano}) for cylindrical nanotubes and by
(\ref{T00D2tor}) and (\ref{TllD2tor}) (with an additional factor 2
which takes into account the presence of two sublattices) for
toroidal nanotubes. The topological Casimir energy is positive for
metallic cylindrical nanotubes and is negative for semiconducting
ones. We have shown that the toroidal compactification of a
cylindrical nanotube along its axis increases the Casimir energy
for periodic boundary conditions and decreases the Casimir energy
for the semiconducting-type compactifications. For finite-length
carbon nanotubes the Casimir forces acting on the tube edges are
always attractive for metallic nanotubes, whereas for
semiconducting-type ones they are attractive for small lengths and
repulsive for large lengths.

\end{document}